\documentclass[showpacs,preprint,amsmath,amssymb]{revtex4}
\usepackage{graphicx}
\usepackage{dcolumn}
\usepackage{bm}

\begin{document}


\title{Plasmon band gap generated by intense ion acoustic waves}

\author{S. Son}
\affiliation{18 Caleb Lane, Princeton, NJ, 08540}
\author{S. Ku}
\email{sku@cims.nyu.edu}
\affiliation{
Courant Institute of Mathematical Sciences, New York University, New York, NY 10012, USA \\
}

\date{\today}

\begin{abstract}
In the presence of an intense ion acoustic wave,  the energy-momentum dispersion relation of plasmons  is strongly modified to exhibit a band gap structure. The intensity of an ion acoustic wave might be measured from the band gap width. The plasmon band gap can be used to block the nonlinear cascading channel of Langmuir wave decay. 
\end{abstract}

\maketitle
The Brillouin scattering  plays an important role in inertial confinement
fusion~\cite{Tabak,Glenzer,Glenzer1} and pulse compression~\cite{Murray}. 
One critical parameter in Brillouin scattering is the ion acoustic wave intensity, $\hat{n}_{\mathrm{iaw}}/\bar{n}_e$,  where $\hat{n}_{\mathrm{iaw}}$ is the perturbed electron density due to the presence of the wave and $\bar{n}_e $ is the equilibrium average electron density. 
At high values of $\hat{n}_{\mathrm{iaw}}/\bar{n}_e$ ($> 0.1$), the ion acoustic wave decays through a few different saturation mechanism, including frequency detuning~\cite{Cohen}, partial wave breaking due to bouncing motion of the trapped particles~\cite{Williams} and nonlinear cascades of ion acoustic waves into other types of waves~\cite{Heikkinen}.
Here, we show that a strong ion acoustic wave modifies the plasmon dispersion relation and generates a plasmon  band gap.  We identify the relevant parameter regime where this gap might be observable.  The gap might  be used to suppress nonlinear cascading of the plasmon decay~\cite{Drake,Heikkinen}.

Electrons in an ion acoustic wave adiabatically follow the electrostatic
potential $\phi_{\mathrm{iaw}}$
and satisfy $n_{\mathrm{iaw}}(x,t) = \bar{n}_{e} \exp( e\phi_{\mathrm{iaw}}(x,t)/T_e) \cong \bar{n}_{e}(1 +e\phi_{\mathrm{iaw}}/T_e )$,
where $T_e$ is the electron temperature.
The Poisson equation is $\nabla^2 \phi_{\mathrm{iaw}} = 4\pi e(n_{\mathrm{iaw}} - n_iZ_i)$, or 
\begin{equation}
(\nabla^2 + 1/\lambda_{\mathrm{de}}^2) \phi_{\mathrm{iaw}} = 4\pi e (\bar{n}_e - Z_i n_i)\mathrm{,}\label{eq:dist}
\end{equation}
where $\lambda_{\mathrm{de}} = ( T_e / 4 \pi \bar{n}_ee^2 )^{1/2} $ is the Debye screening length.
The closure of the above equation, considering the momentum and continuity equations for ions,
results in a dispersion relation 
\begin{equation}
\omega_i(k) = \frac{v_s k}{\sqrt{1 + (k \lambda_{\mathrm{de}})^2}}\label{eq:dis}\mathrm{,}
\end{equation}
where $v_s = (Z_iT_e/m_i)^{1/2}$ is the ion acoustic wave velocity.
The electron density, in the presence of an ion acoustic wave, would exhibit spatial periodicity: $n_{\mathrm{iaw}} = \bar{n}_e ( 1 + e\hat\phi_{\mathrm{iaw}}/T_e\sin(kx-\omega_i t))$, where $\omega_i$ and $k$ satisfy Eq.~(\ref{eq:dis}) so that $ e\hat\phi_{\mathrm{iaw}}/T_e = \hat{n}_{\mathrm{iaw}}/\bar{n}_e$, and $\hat\phi_{\mathrm{iaw}}$ is the amplitude of ion acoustic wave. 
The ratio between the Langmuir wave frequency $\omega_{\mathrm{pe}} = (4 \pi \bar{n}_e e^2/m_e)^{1/2}$ and $\omega_i$ in Eq.~(\ref{eq:dis}) gives

\begin{equation}
\frac{\omega_i}{\omega_{\mathrm{pe}}}  =\sqrt{\frac{Z_i m_e}{m_i}} \frac{k\lambda_{\mathrm{de}}}{\sqrt{1 + ( k\lambda_{\mathrm{de} })^2}} \mathrm{.} 
\end{equation}
Since $k\lambda_{\mathrm{de}}<1$ in the cases we consider,
$\omega_i/\omega_{\mathrm{pe}}\ll 1$ is satisfied and the electron fluid equations 
can be approximately solved by fixing the ion density $n_i$ and  the potential $\phi_{\mathrm{iaw}} $. 
The fluid equations together with the Poisson equation for the electrons are given by
\begin{eqnarray}
\frac{\partial n_e}{\partial t} + \nabla\cdot (n_e \mathbf{v_e}) = 0 \mathrm{,} \nonumber \\ \nonumber \\
\frac{\partial \mathbf{v_e}}{\partial t} = \frac{e}{m_e}\mathbf{\nabla} \phi -\frac{1}{m_en_e(x,t)} \mathbf{\nabla} p_e\nonumber  \mathrm{,}\\ \nonumber \\
\nabla^2 \phi =  4\pi e(n_e -Z_i n_i)  \mathrm{,}\nonumber \\ \nonumber 
\end{eqnarray}
where $\mathbf{v_e}$ is the electron fluid velocity and $p_e$ is the electron pressure.

The above equations can be linearized by $\mathbf{v_e} = \mathbf{v_1} $, $n_e = n_{\mathrm{iaw}}(x) + n_1$, $\phi = \phi_{\mathrm{iaw}} + \phi_1 $, and $p_e =  p_{\mathrm{iaw}} + p_1$:
\begin{eqnarray}
\frac{\partial n_1}{\partial t} + \nabla\cdot (n_{\mathrm{iaw}}(x)\mathbf{v_1}) = 0 \label{eq:cont} \mathrm{,}\\ \nonumber \\
\frac{\partial \mathbf{v_1}}{\partial t}=  \frac{e}{m_e}\mathbf{\nabla} \phi_1 -\frac{\gamma~T_e}{m_e n_{\mathrm{iaw}}(x)} \mathbf{\nabla} n_1 + \frac{T_e \nabla n_\mathrm{iaw}(x)  }{m_e n_\mathrm{iaw}^2(x)} n_1 \mathrm{,} \label{eq:mom} \\  \nonumber \\
\nabla^2 \phi_1 =  4\pi e n_1   \label{eq:pos} \mathrm{,}\\ \nonumber
\end{eqnarray}
where ion motion is ignored and $\nabla p_1 = \gamma T_e \nabla n_1$ and $\nabla p_\mathrm{iaw} = T_e \nabla n_\mathrm{iaw}$ are assumed. $\gamma$ is the heat capacity ratio.
Note that $n_{\mathrm{iaw}}$, $p_{\mathrm{iaw}}$ and $\phi_{\mathrm{iaw}}$ already satisfy the Poisson equation and the momentum equation 
from the ion acoustic wave: $\nabla^2 \phi_{\mathrm{iaw}} = 4 \pi e(n_{\mathrm{iaw}} - n_iZ_i)$ and $\nabla p_\mathrm{iaw} = e n_\mathrm{iaw} \nabla \phi_\mathrm{iaw}$.
By taking the time derivative of Eq.~(\ref{eq:cont}), and substituting  Eqs.~(\ref{eq:mom}) and (\ref{eq:pos}) into Eq.~(\ref{eq:cont}), we obtain

\begin{figure}
\scalebox{0.6}{
\includegraphics{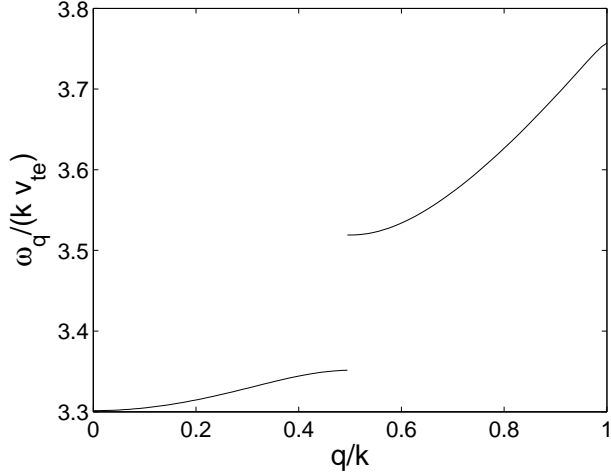}}
\caption{\label{fig:1}
The dispersion relation for the Langmuir wave when $k\lambda_{\mathrm{de}}=0.3$,
$e\phi_{\mathrm{iaw,0}/}T_e = 0.1$, $Z_i=1$, and $\gamma = 3$.}
\end{figure}
\begin{figure}
\scalebox{0.6}{
\includegraphics{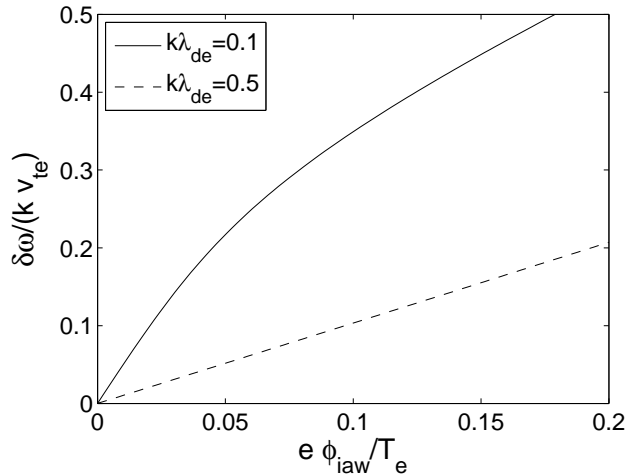}}
\caption{\label{fig:2}
The gap width $\delta \omega$ as a function of $e\hat\phi_{\mathrm{iaw}}/T_e$ for the two cases
$k\lambda_{\mathrm{de}}=0.1$ and $0.5$. $Z_i=1$ and $\gamma = 3$.}.
\end{figure}


\begin{equation} 
\frac{\partial^2 \phi_{1x}}{\partial t^2} =  -\omega^2_{\mathrm{pe}}(x)  \phi_{1x} + \frac{\gamma T_e}{m_e} \nabla^2\phi_{1x} - \frac{T_e}{m_e}\frac{\nabla n_\mathrm{iaw}}{n_\mathrm{iaw}} \nabla \phi_{1x}  \label{eq:band} \mathrm{,}
\end{equation}
where $\omega^2_{\mathrm{pe}}(x)  = 4\pi n_{\mathrm{iaw}}(x) e^2/m_e$, $\phi_{1x} = d\phi_1 / dx$, and only $x$-dependency is assumed. 
Note that the first and third terms on the right hand side of Eq.(\ref{eq:band}) have a periodicity
since $n_{\mathrm{iaw}}(x) = n_{\mathrm{iaw}}(x+2\pi/k)$.  Using Bloch's theorem,
we could solve $ \phi_q(x) = \exp(iqx) u_q(x)$ for a given $q$, where $u_q(x) = u_q(x+2\pi/k)$. 
A similar equations have been used to obtain the dispersion relation in the photonic band gap material~\cite{Yablo}.
We obtain the dispersion relation $ \omega(q) $ for $k\lambda_{\mathrm{de}}=0.3$, $e\phi_{\mathrm{iaw,0}}/T_e = 0.1$ with $Z_i=1$ and $\gamma =3$ (Fig.~\ref{fig:1}).  
The plasmon dispersion deviates considerably from the conventional one; there exists
a band where $ q =\pm k/2$.
The band gap size $\delta \omega = \omega(k/2+) -\omega(k/2-)$ varies with $e\hat\phi_{\mathrm{iaw}}/T_e$. 
The gap size $\delta \omega$ is calculated as a function of $e\hat\phi_{\mathrm{iaw}}/T_e$
when $k\lambda_{\mathrm{de}}=0.1$ and $0.5$ (Fig.~\ref{fig:2}).
The larger the ion acoustic wave intensity $e\hat\phi_{\mathrm{iaw}}/T_e$, the bigger the band gap size $\delta \omega $. The gap, which may be  measured from the frequency resolved with Thomson scattering~\cite{Chihara, Gregori,Glenzer,Baker},  has information about $e\hat\phi_{\mathrm{iaw}}/T_e$.

However, a few conditions need to be met for a measurable plasmon band to exist.
First, for a given $k$, there should be a meaningful collective Langmuir wave.
This imposes the condition $q \cong k/2 <k_F$ for a degenerate plasma,
where $k_F$ is the Fermi wave vector and $q < 0.53 \lambda_{\mathrm{de}}^{-1}$ for a hot plasma. 
Second, the electron temperature should be high enough so that the plasmon is
thermally excited ($T_e > \hbar \omega_{\mathrm{pe}}$) as well.  
Third, the plasmon width, roughly given as
$(1/2)\omega_{\mathrm{pe}} \mathrm{Im}(\epsilon(q,\omega))$ when $\mathrm{Re}(\epsilon(q,\omega)) = 0$,
should be smaller than the gap size.
Assuming $e\hat\phi_{\mathrm{iaw}}/T_e \ll 1$, the gap size from a perturbation analysis is estimated to be $(\omega_{\mathrm{pe}}^2/ qv_{\mathrm{te}})(e\hat\phi_{\mathrm{iaw}}/T_e)$, and the condition becomes $\mathrm{Im}(\epsilon(q,\omega)) < 2 (\omega_{\mathrm{pe}}/ qv_{\mathrm{te}})(e\hat\phi_{\mathrm{iaw}}/T_e)$.  
This condition would be satisfied either by having a large $e\hat\phi_{\mathrm{iaw}}/T_e$ or by decreasing $k$ so that $\omega/k \gg v_{\mathrm{te}}$.  For a  classical hot plasma, the condition can be estimated as 
\begin{equation}
 \frac{(\pi/2)^{1/2}}{(k \lambda_{\mathrm{de}})^2} \exp \left( -\frac{1}{2} \frac{1}{(k \lambda_{\mathrm{de}})^2}\right) < 2 \frac{e\hat\phi_{\mathrm{iaw}}}{T_e} \mathrm{.}
\end{equation}
When $e\hat\phi_{\mathrm{iaw}}/T_e = 0.1 $, it would be satisfied if $k \lambda_{\mathrm{de}} < 1/ 3$.   Lastly, a perfectly monochromatic ion wave is a somewhat artificial circumstance in nature or a laboratory. 
If the spectrum of the ion acoustic waves is peaked but has finite spectral width,  the treatment here could be irrelevant.   For a given relative spectral width, $\delta f/ f $,  the quasi-periodic density structure is coherent over a length $L$ given roughly by as $ L =  2\pi /q  (f/\delta f)$. The decay length of the forbidden plasmon band, which is the so-called skin depth of the transient state, is roughly given as $ L_{\mathrm{skin}} = v_{\mathrm{te}}/ \delta \omega $.
For the plasmon band to be relevant, the periodic length, $L$, should be larger than the skin depth, $  L_{\mathrm{skin}}  $: 

\begin{equation}
  \frac{\delta f }{ f} < \frac{2\pi \delta \omega }{q v_{\mathrm{te}}} \mathrm{.}
\end{equation}
For a weak band gap where the perturbation analysis is relevant,  this can be estimated as $ \delta f / f <  2 \pi  (\omega_{\mathrm{pe}}^2/ (qv_{\mathrm{te}})^2 )(e\hat\phi_{\mathrm{iaw}}/T_e)$. This condition is relevant to various experiments.

\begin{figure}
\scalebox{0.6}{
\includegraphics{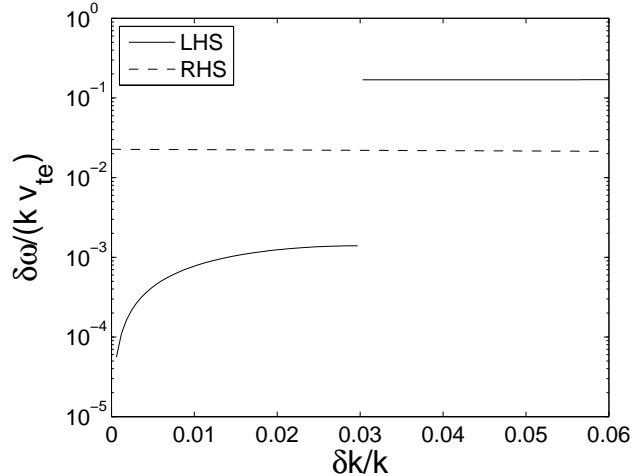}}
\caption{\label{fig:4}
The right hand side and the left hand side of Eq.~(\ref{eq:jmp}),
where $k\lambda_{\mathrm{\mathrm{de}}} = 0.3$, $e\phi_{\mathrm{iaw}} = 0.1 $, $Z_i = 1$, $m_i=1836 m_e$, $\gamma=3$, and $k_1 = 0.53k$.}
\end{figure}

In backward Raman scattering \cite{Murray, McKinstrie, McKinstrie2}, there exists an intensity threshold  for  nonlinear saturation of  Langmuir waves \cite{Drake,Heikkinen,Kirkwood}. One of the main decay mechanism is the cascading of a plasmon into an ion acoustic wave and another plasmon. As shown in Fig.~\ref{fig:1}, an intense ion acoustic wave generates a very strong nonlinear dispersion relation for the plasmon, which might be used to block some of the decay channel.  Consider a situation where
a Langmuir wave with $k_1$ decays into a Langmuir with the wave vector $k_2=k_1 -\delta k$ and an ion acoustic wave with the wave vector $k_1 + k_2=2k_1 - \delta k$, as studied in Ref.~\cite{Heikkinen}.
The energy conservation condition between the waves  is

\begin{equation}
 \omega(k_1) -\omega(k_1-\delta k) = \frac{(2k_1-\delta k)v_s }{1 + (k\lambda_{\mathrm{de}})^2} \label{eq:jmp} \mathrm{.}
\end{equation}  
If $k_1$ is slightly larger than $k/2$, then there will be a jump in the right hand side of Eq.~(\ref{eq:jmp}) around $k_1 - \delta k \cong k/2$ (Fig.~(\ref{fig:4})).
If $(2k_1v_s)/ (1 + (k\lambda_{\mathrm{de}})^2) < \delta \omega$, there is no $\delta k$ satisfying Eq.~(\ref{eq:jmp}), as shown in Fig.~\ref{fig:4}.  This fact could be used to block some parasitic nonlinear cascading of Langmuir waves in backward Raman scattering \cite{Drake, Heikkinen,Kirkwood}. 
It has been shown previously  by Barr and Chen \cite{chen} that stimulated Raman scattering could be suppressed in the presence of the periodic electron density structure. 
The difference between their result  and ours is the fact that the wave vector here is $q/2$ instead of $q$,  and the plasmon decay into other plasmons is considered instead of stimulated Raman scattering.
Whether the work of Barr and Chen is relevant to our result would be an interesting question. 
 In previous work \cite{kruer, dawson}, it has been shown that the periodic electron density could be facilitated by the nonlinear  decay of the long wave length plasmon into a short wave length plasmon.  
Here the decay is from a plasmon to a plasmon with a comparable wave length. An important question would be whether the suppression effect discussed here can overcome the decay channel described by Kruer~\cite{kruer}.  
The answer might depend on the physical parameters.  
For an example, consider a very dense plasma.  
As shown in Ref.~\cite{landau}, Landau damping could be very intense, but the plasmon could be excited to a considerable intensity without damping. 
If this is the case, the plasmon decay described by Kruer is very weak because of the heavy damping of the excited wave whereas the original plasmon could be very strong.  The band gap is then a very efficient method to suppress the plasmon decay.  

Here, we showed that a strong ion acoustic wave  generates a plasmon band gap structure and that the ion acoustic wave intensity could be determined from the gap size. 
We also showed that some of the Langmuir wave cascading channel could be blocked using the generated band gap.  In the presence of the band gap, the Langmuir wave would exhibit exotic properties as an electron does in a solid lattice potential. The group velocity could be almost zero around the band gap and the plasmon could get even reflected, stretched or compressed~\cite{Fisch}. The plasmon band gap suggested in this paper would be useful in laser interaction with dense plasmas along with recently discovered new physics \cite{landau, sonprl, sonpla}.   

We would like to thank Dr. S. J. Moon, Dr. D. Clark, and Prof. N. Fisch for carefully reading the manuscript and for useful discussions and advice.

\end{document}